\documentclass{aa}
\usepackage{txfonts}
\usepackage{times}
\usepackage{natbib}
\usepackage{graphicx}
\usepackage{epsfig,longtable,lscape}
\usepackage{epsfig}
\usepackage{rotating}
\usepackage{hyperref}
\bibpunct[]{(}{)}{;}{a}{}{,}

\def\XMM{{\em XMM--Newton}}
\def\ROSAT{{\em ROSAT}}
\def\Swift{{\em Swift}}
\def\Chandra{{\em Chandra}}
\def\pn{{\em pn}}

\def\cxo{{CXOU\,J225355.1+624336}}

\def\approxgt{\mathrel{\hbox{\rlap{\lower.55ex \hbox {$\sim$}}
        \kern-.3em \raise.4ex \hbox{$>$}}}}
\def\approxlt{\mathrel{\hbox{\rlap{\lower.55ex \hbox {$\sim$}}
        \kern-.3em \raise.4ex \hbox{$<$}}}}
\def\pdot{\dot P}

\def\lx{$L_{\rm X}$}
\def\fx{$f_{\rm X}$}
\def\flux{\mbox{erg cm$^{-2}$ s$^{-1}$}}
\def\lum{\mbox{erg s$^{-1}$}}
\def\countsec{\hbox{counts s$^{-1}$}}
\def\fph{ph cm$^{-2}$ s$^{-1}$}
\def\nh{$N_{\rm H}$}
\def\ltsima{$\; \buildrel < \over \sim \;$}
\def\lsim{\lower.5ex\hbox{\ltsima}}
\def\gtsima{$\; \buildrel > \over \sim \;$}
\def\gsim{\lower.5ex\hbox{\gtsima}}

\def\chisqnu {$\chi^{2}_{\nu}$}

\begin{document}

\title{A deep \XMM\ observation of the X-Persei-like binary system \cxo\thanks{Based on observations obtained with \XMM, an ESA science mission, with instruments and contributions directly funded by ESA Member States and NASA}}

\author{N.~La~Palombara\inst{1}, L.~Sidoli\inst{1}, P.~Esposito\inst{1,2}, G. L. Israel\inst{3}, G. A. Rodr\'iguez Castillo\inst{4}}

\institute{INAF--Istituto di Astrofisica Spaziale e Fisica Cosmica di Milano, 
      via A. Corti 12, 20133 Milano, Italy\\
              e-mail: \href{mailto:nicola.lapalombara@inaf.it}{nicola.lapalombara@inaf.it}
                \and Scuola Universitaria Superiore IUSS Pavia, Palazzo del Broletto, piazza della Vittoria 15, 27100 Pavia, Italy
             \and INAF--Osservatorio Astronomico di Roma, via Frascati 33, 00040 Monteporzio Catone, Italy
             \and INAF -- Istituto di Astrofisica Spaziale e Fisica Cosmica di Palermo, Via U. La Malfa 153, 90146 Palermo, Italy}
               
               \date{Received DD Month YYYY; accepted DD Month YYYY}

\titlerunning{\XMM\ observation of \cxo}

\authorrunning{N.~La~Palombara et al.}

\abstract{We report on the follow-up \XMM\ observation of the persistent X-ray pulsar \cxo, which was discovered with the CATS@BAR project on archival \Chandra\ data. The source was detected at \fx(0.5-10 keV) = 3.4$\times 10^{-12}$ \flux, a flux level that is fully consistent with previous observations performed with \ROSAT, \Swift, and \Chandra. When compared with previous measurements, the measured pulse period $P$ = 46.753(3) s implies a constant spin down at an average rate of $\pdot = 5.3\times 10^{-10}$ s s$^{-1}$. The pulse profile is energy dependent, showing three peaks at low energy and a less structured profile above about 3.5 keV. The pulsed fraction slightly increases with energy. We described the time-averaged EPIC spectrum with four different emission models: a partially covered power law, a cutoff power law, and a power law with an additional thermal component (either a black body or a collisionally ionised gas). In all cases we obtained equally good fits, so it was not possible to prefer or reject any emission model on a statistical basis. However, we disfavour the presence of thermal components since their modeled X-ray flux, resulting from a region larger than the neutron star surface, would largely dominate the X-ray emission from the pulsar. The phase-resolved spectral analysis showed that a simple flux variation cannot explain the source variability and proved that there is a spectral variability along the pulse phase. The results of the \XMM\ observation confirmed that \cxo\ is a Be X-ray binary (BeXB) with a low luminosity (\lx $\sim 10^{34-35}$ \lum), limited variability, and a constant spin down. Therefore, these results reinforce its source classification as a persistent BeXB.}


\maketitle

\section{Introduction}   

High-mass X-ray binaries (HMXBs; \citealt{Kretschmar2019}) can be broadly divided into three categories: supergiant binaries, Be X-ray binaries (BeXBs), and the emerging group of supergiant fast X-ray transients \citep{Sidoli2017}, the third of which shares properties with the two other classes \citep[e.g.][]{Enoto2019}. Most BeXBs host a pulsating neutron star (NS) in a wide and highly eccentric orbit ($e > 0.3$) and are bright (X-ray luminosity $L_{\mathrm{X}}>10^{36}$\,\lum) transient systems. The transient X-ray emission can be ascribed either to enhanced accretion from the circumstellar decretion disc of the Be donor as the NS approaches the periastron (periodic `type I' outbursts, lasting for $\approx$20--30\% of the orbital period) or to unpredictable episodes of mass loss from the Be star (more luminous `type II' or `giant'  outbursts, which can last from weeks to months; e.g. \citealt{Reig2011}). However, a particularly relevant sub-class of BeXBs are the persistent ones \citep{Pfahl2002} as they are suspected to contribute to (at least) a few percent  of the unidentified Galactic X-ray sources with luminosity $L_{\mathrm{X}}\approx10^{35}$\,\lum. These sources, the prototype of which is X Persei, produce such a comparatively dim luminosity by steadily accreting the wind material in wide (orbital periods longer than $\sim$30\,d) and nearly circular ($e < 0.2$) orbits. 

\citet{Esposito2013} individuated a candidate member of the group in the X-ray source \cxo\ based on: (i) the discovery of the NS spin period \citep[47\,s; see also][]{Israel2016}, (ii) the moderate luminosity ($L_\mathrm{X}<10^{34}$\,erg\,s$^{-1}$), which is seemingly steady over X-ray observations spanning $\sim$16\,yr, and (iii) the optical follow-up, which suggests a Be companion in the Perseus arm of the Galaxy. In this work, we report on the results from the analysis of a deep \XMM\ exposure that was performed about 5\,yr after the observations discussed in \citet{Esposito2013}, stretching the X-ray history of the source to more than two decades.

         %%%%%%%%%%%%%%%%%%%%%%%%%%%%%%%%%%%%%%%%%%%%%%%%%%%%%%%%%%%%%%%%%%%%
         \section{Observation and data reduction }
         \label{data}
         %%%%%%%%%%%%%%%%%%%%%%%%%%%%%%%%%%%%%%%%%%%%%%%%%%%%%%%%%%%%%%%%%%%%

\cxo\ was observed with \XMM\ on February 28, 2014 (MJD 56716). The total observing time was $\simeq$ 30 ks. In Table~\ref{observation} we report the setups, time resolutions, and net exposure times of the three EPIC cameras -- two MOS \citep{Turner+01} and one \pn\ \citep{Struder+01} -- and of the Reflection Grating Spectrometer (RGS; \citealt{denHerder+01}). We processed the events with version 18 of \XMM~{\em Science Analysis System}\footnote{https://xmm-tools.cosmos.esa.int/external/xmm\_user\_support/documentation/sas\_usg/USG/} (\texttt{SAS}). We verified that $\simeq$ 3.5 ks of the observation were affected by soft-proton contamination, which could jeopardise the results of the EPIC data analysis (while its impact on the RGS data is negligible). Therefore, for the EPIC data we ignored the time intervals affected by this contamination and only considered the remaining parts of the observation. In Table~\ref{observation} we report the effective exposure time of each instrument, which, for the EPIC cameras, takes into account the dead times of 13\,\% and 2.5\,\% for \pn\ and MOS, respectively. Due to the very low count rate (CR), we did not use the RGS data.

For the analysis of the EPIC data, we considered events with patterns between 0 and 12 (covering up to four nearby pixels) for the MOS cameras and events with patterns between 0 and 4 (corresponding to mono- and bi-pixel events) for the \pn\ camera. Since the source was rather faint, for the event selection we considered a relatively small extraction radius of 30 arcsec around the source position. As such, we maximised the source contribution and minimised that of the background events; these were selected from a circular region offset from the target position and free of sources, with an extraction radius of 120 arcsec. The CR of the source was rather low (Table~\ref{observation}), and we verified that for all the EPIC cameras the data were not affected by photon pileup.

\begin{table*}
\caption{Summary of the \XMM\ observation of \cxo\ (ID 0743980301).}\label{observation}
\vspace{-1 cm}
\begin{center}
\begin{tabular}{ccccccc} \\ \hline
Instrument      & Filter        & Mode                  & Time Resolution 
      & Net Exposure Time       & Extraction Radius     & Net Count Rate  
\\
                &               &                       &                 
      & (ks)                    & (arcsec)              & (\countsec)     
\\ \hline
\pn\            & Thin 1        & Full Frame            & 73 ms           
       & 20.4                    & 30                    & 0.43$\pm$0.03  
 \\
MOS1            & Thin 1        & Full Frame            & 2.7 s           
      & 25.1                    & 30                    & 0.134$\pm$0.002 
  \\
MOS2            & Thin 1        & Full Frame            & 2.7 s           
      & 25.2                    & 30                    & 0.131$\pm$0.002 
  \\
RGS1            & -             & Spectroscopy          & 4.8 s           
      & 29.0                    & -                     & - \\
RGS2            & -             & Spectroscopy          & 9.6 s           
      & 29.0                    & -                     & - \\ \hline
\end{tabular}
\end{center}
\end{table*}

         %%%%%%%%%%%%%%%%%%%%%%%%%%%%%%%%%%%%%%%%%%%%%%%%%%%%%%%%%%%%%%%%%%%%
         \section{Timing analysis}
         \label{timing}
         %%%%%%%%%%%%%%%%%%%%%%%%%%%%%%%%%%%%%%%%%%%%%%%%%%%%%%%%%%%%%%%%%%%%

\begin{figure}
\begin{center}
\includegraphics[width=8.5cm,angle=0]{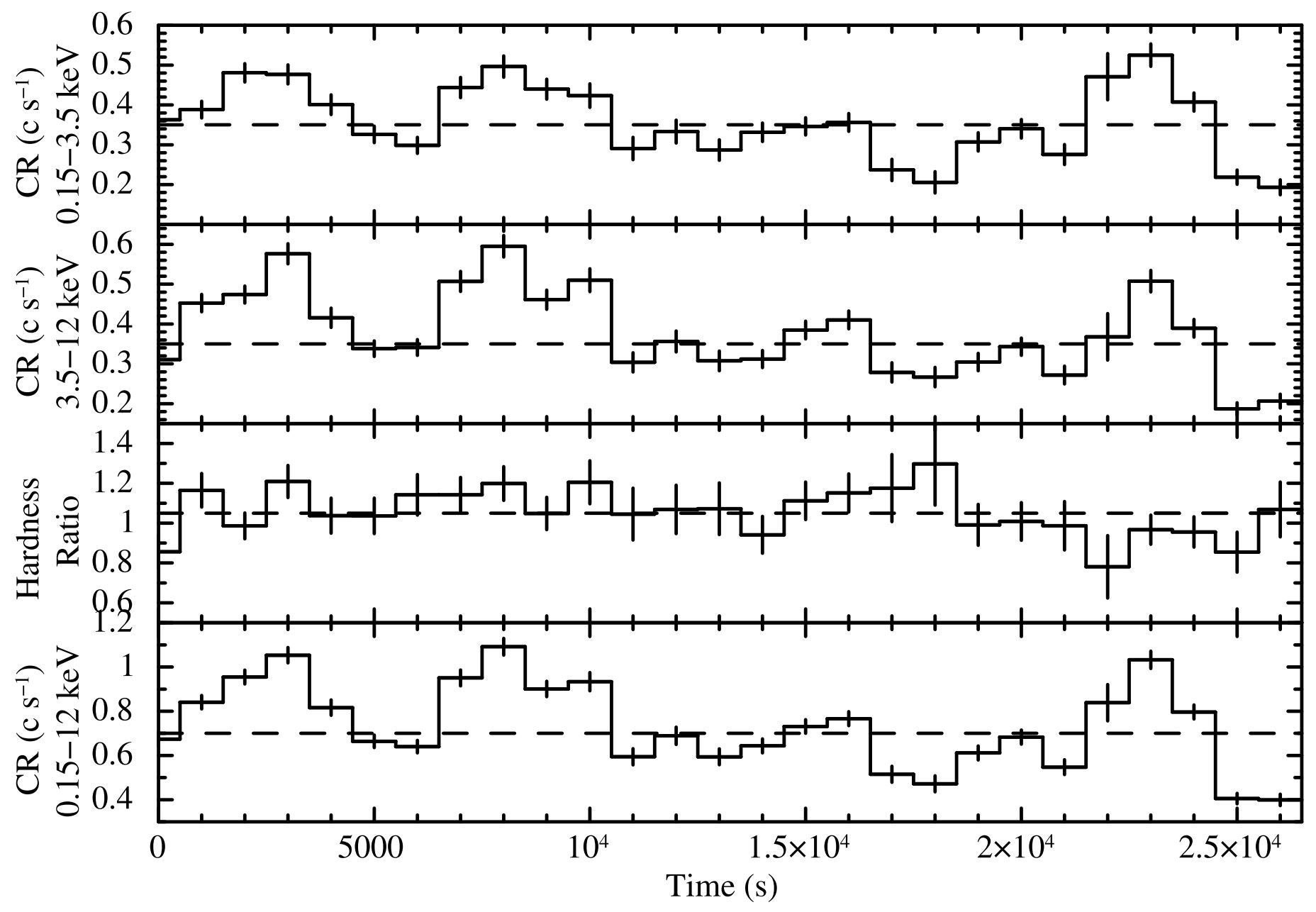}
\caption{Background-subtracted light curves of \cxo, corrected for the extraction region, in the energy ranges 0.15--3.5, 3.5--12, and 0.15--12 keV with a time binning of 1000 s. The horizontal dashed lines represent the average values.}
\label{lc}
\end{center}
\vspace{-0.75 cm}
\end{figure}

For the timing analysis of the EPIC events, we used the \textsc{sas} tool \textsc{barycenter} to report their arrival times to the Solar System barycentre. In addition to the total energy range between 0.15 and 12 keV, we also considered the two energy ranges 0.15-3.5 keV (soft) and 3.5-12 keV (hard), which include a similar number of source counts. Then, we built a light curve (with a time binning of 1000 s) for each of the three ranges and each of the three cameras. To this aim, we used the \textsc{sas} tool \textsc{epiclccorr} to correct for the background and the extraction region. The average CR in the total range was $\simeq$ 0.43 cts s$^{-1}$ in the \pn\ camera and $\simeq$  0.13 cts s$^{-1}$ in each of the two MOS cameras. For each energy range, we summed the light curves of the individual cameras to obtain the cumulative light curve. In Fig.~\ref{lc} the three curves and the hardness ratio (HR) of the hard (H) to the soft (S) light curves (HR = H/S) are shown. The average CR was $\simeq$ 0.35 cts s$^{-1}$ in both the soft and hard energy ranges. The light curve shows the high source variability over short timescales: the CR even varies by a factor $\sim$ 2 between consecutive time bins. However, the source flux does not show any increasing or decreasing trend over the observation timescale. On the other hand, the HR does not vary significantly. It remains almost constant around 1 and does not depend on the source CR.

%%%%%%%%%%%%%%%%%%%%%%%%%%%%%%%%%%%%%%%%%%%%%%%%%%%%%%%%%%%%%%%%%%%%%%%% 
%%%%%%%%%%%%%%%%%%%%%%%%%%%%%%%%%%%%%%%%%%%%%%%%%%%%%%%%%%%%%%%%%%%%%%%%

\begin{figure}
\begin{center}
\includegraphics[width=8.5cm,angle=0]{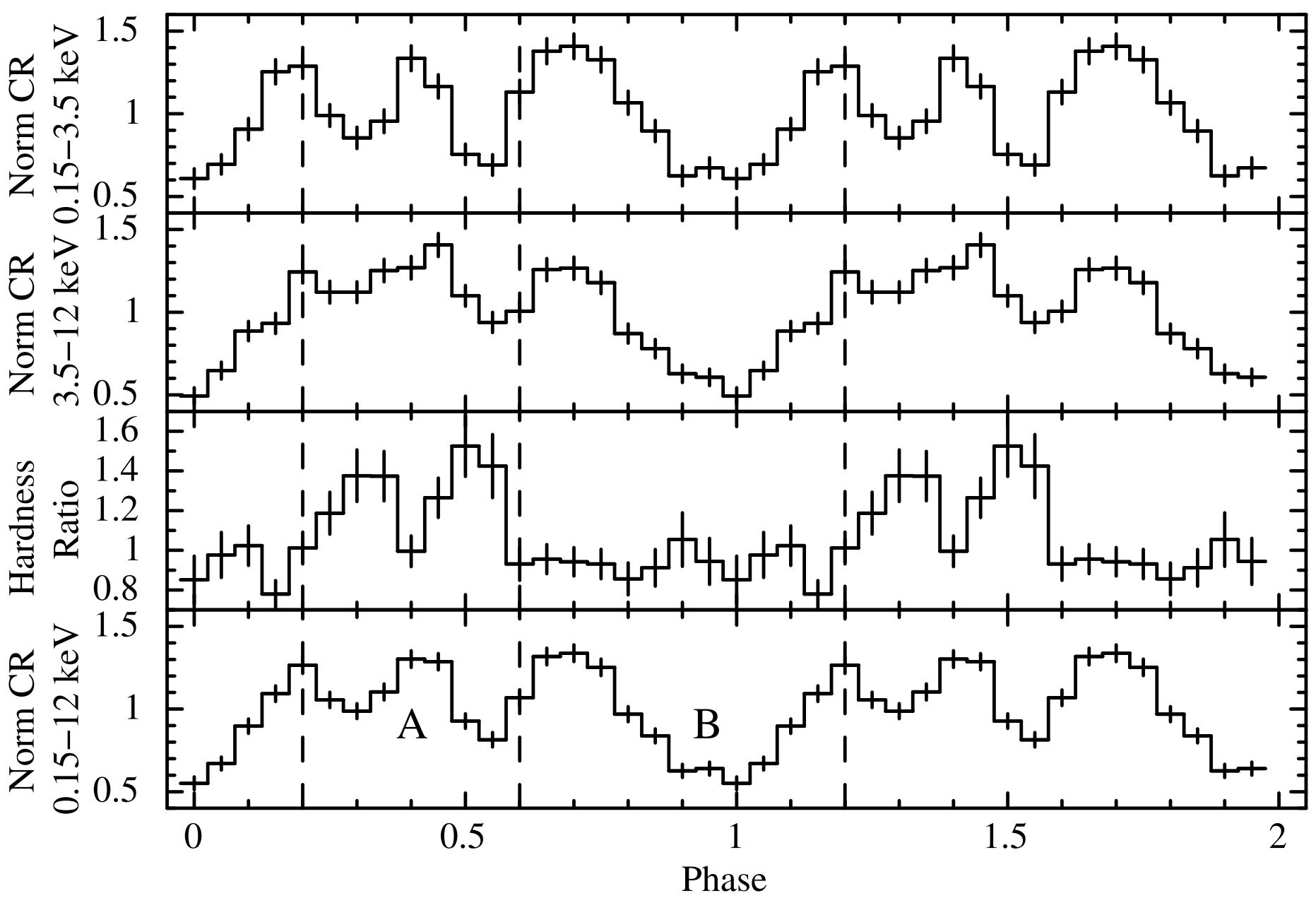}
\caption{Pulse profile and HR of \cxo\ in the energy ranges 0.15--3.5, 3.5--12, and 0.15--12 keV. The dashed vertical lines delimit the two phase intervals, A and B (corresponding, respectively, to the maximum and minimum of the HR), selected for the phase-resolved spectral analysis.}
\label{flc2E}
\end{center}
\vspace{-0.75 cm}
\end{figure}

For the measurement of the pulse period, we merged the event datasets of the three instruments into a unique dataset in order to increase the count statistics. Then, we applied a standard phase-fitting technique, thus obtaining a best-fitting period of $P$ = 46.753(3) s. In Fig.~\ref{flc2E} we report the normalised folded light curve in the three energy ranges, together with the HR between the hard and soft curves. The shape of the pulse profile is rather similar in the two energy ranges; in both cases it shows three main peaks at the pulse phases $\Phi \simeq$ 0.2, 0.4, and 0.7. However, the first and second peaks are more evident and distinct in the soft range, where they are separated by a deeper minimum than in the hard range. We also note that the HR is rather anti-correlated with the source CR since it shows an increase in coincidence with the two CR minima between the CR peaks (at $\Phi \simeq$ 0.3 and 0.5, respectively).

%%%%%%%%%%%%%%%%%%%%%%%%%%%%%%%%%%%%%%%%%%%%%%%%%%%%%%%%%%%%%%%%%%%%%%%% 
%%%%%%%%%%%%%%%%%%%%%%%%%%%%%%%%%%%%%%%%%%%%%%%%%%%%%%%%%%%%%%%%%%%%%%%%

\begin{figure}
\begin{center}
\includegraphics[width=8.5cm,angle=0]{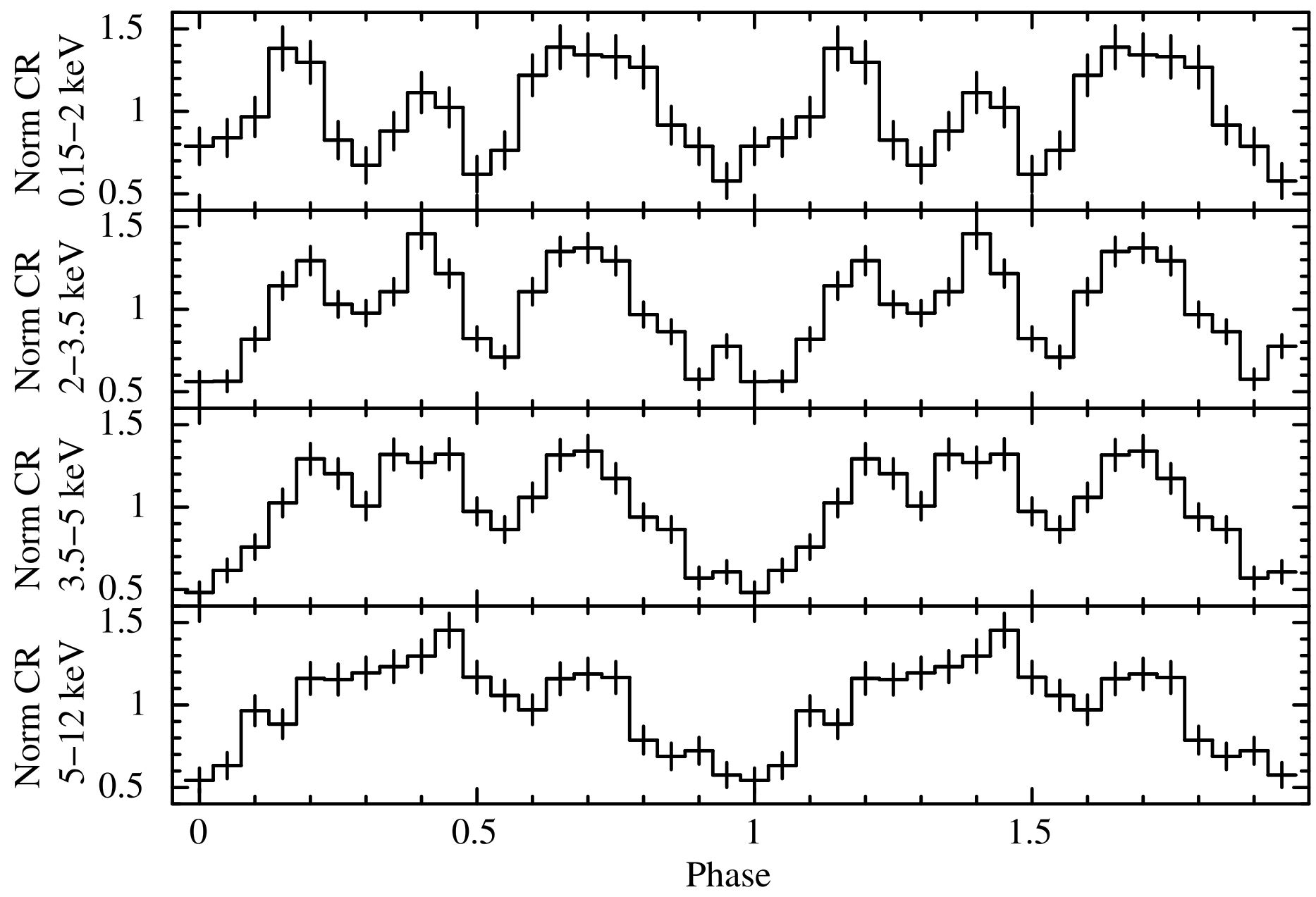}
\caption{Pulse profile of \cxo\ in the energy ranges 0.15--2, 2--3.5, 3.5--5, and 5--12 keV.}
\label{flc4E}
\end{center}
\vspace{-0.75 cm}
\end{figure}

In order to better study the energy dependence of the pulse profile, we also considered four narrower energy bands. In Fig.~\ref{flc4E} we show the folded light curve in the energy ranges 0.15--2, 2--3.5, 3.5--5, and 5--12 keV. It shows clearly that the pulse profile is strongly energy dependent. The first peak, at $\Phi \simeq$ 0.2, is prominent at E $<$ 2 keV and decreases at higher energies. The second peak, at $\Phi \simeq$ 0.4, becomes more prominent at high energies, while the third peak, at $\Phi \simeq$ 0.7, tends to decrease. The flux variability is almost constant with energy. In fact, the average pulsed fraction (PF), defined as PF = (CR$_{\rm max}$ - CR$_{\rm min}$)/(2$\times$CR$_{\rm average}$), increases from $\simeq$ 40 \% at E $<$ 2 keV to $\simeq$ 45 \% at E $>$ 5 keV.

%%%%%%%%%%%%%%%%%%%%%%%%%%%%%%%%%%%%%%%%%%%%%%%%%%%%%%%%%%%%%%%%%%%%%%%% 
%%%%%%%%%%%%%%%%%%%%%%%%%%%%%%%%%%%%%%%%%%%%%%%%%%%%%%%%%%%%%%%%%%%%%%%%

         %%%%%%%%%%%%%%%%%%%%%%%%%%%%%%%%%%%%%%%%%%%%%%%%%%%%%%%%%%%%%%%%%%%%
         \section{EPIC time-averaged spectral analysis}
         \label{average-spectroscopy}
         %%%%%%%%%%%%%%%%%%%%%%%%%%%%%%%%%%%%%%%%%%%%%%%%%%%%%%%%%%%%%%%%%%%%

The light curve reported in Fig.~\ref{lc} shows that, during the \XMM\ observation of \cxo, the spectral properties of the target did not change significantly; its HR remained almost constant. Even if the source flux is variable on a short timescale, it does not vary over the observation timescale. Therefore, we performed the source spectral analysis by accumulating, for each EPIC camera, the source spectrum over the full exposure; for this purpose, the source events were extracted from the same regions we used for the light curves. Each of the three spectra was re-binned with a significance of at least 3 $\sigma$ for each energy bin, using the \texttt{SAS} tool \textsc{specgroup}. The spectral binning was larger than the intrinsic spectral resolution of the instrument; moreover, we removed the spectral channels that, after the background subtraction, were consistent with zero. We generated the corresponding response matrices and ancillary files using the \texttt{SAS} tasks \textsc{rmfgen} and \textsc{arfgen}, respectively. Since there were no significant bins at lower and higher energies, we considered the energy range 0.5-10 keV for the spectral analysis, and we used version 12.9.1 of \texttt{xspec}. We calculated the spectral uncertainties at the 90 \% confidence level for each interesting parameter. For the source distance we assumed a value of 5 kpc. Since we verified that the separate fits of the three EPIC spectra provided consistent results, we fitted them simultaneously. We took into account possible uncertainties in the instrumental responses by leaving the relative normalisations of the three cameras free to vary. We fixed the normalisation factor of the \pn\ spectrum to 1 and found that the relative normalisation factors for the MOS spectra were 0.83 $\pm$ 0.02 for MOS1 and 0.97 $\pm$ 0.02 for MOS2. For the spectral fitting, we used the elemental abundances provided by \citet{WilmsAllenMcCray00}, the photoelectric absorption cross-sections of \citet{Verner+96}, and the absorption model \textsc{tbabs} in \texttt{XSPEC}. The estimated fluxes were calculated with the \texttt{XSPEC} tool \textsc{cflux} in the energy range 0.5-10 keV.

\begin{figure}
\begin{center}
\includegraphics[width=8.5cm,angle=0]{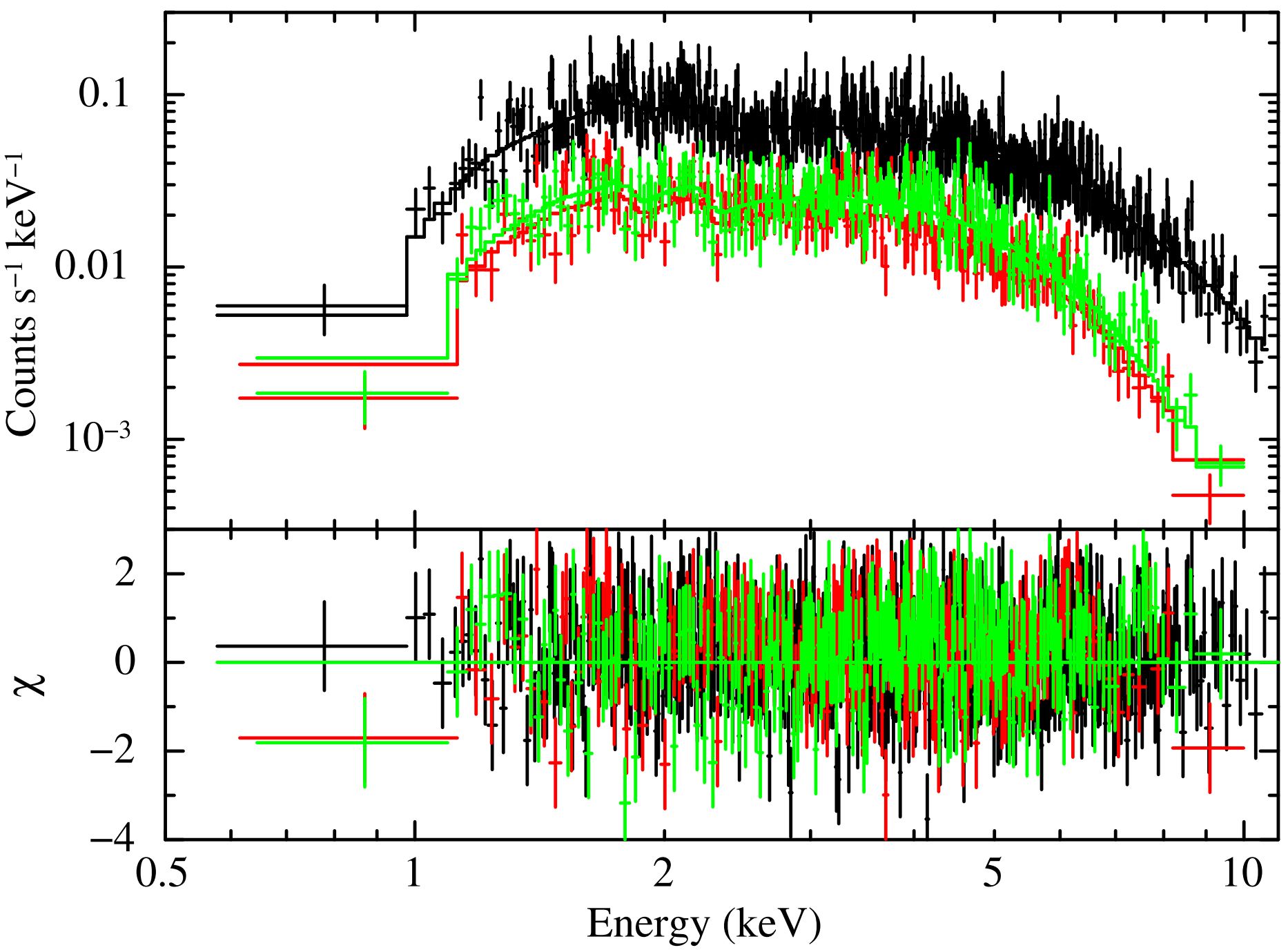}
\caption{Time-averaged spectrum of \cxo. The \pn, MOS1, and MOS2 data are 
represented, respectively, with black, red, and green symbols. \textit{Upper panel}: Superposition of the EPIC spectra with the best-fitting TBPCF$\times$PL emission model. \textit{Lower panel}: Data-model residuals in units of standard deviation ($\sigma$).}
\label{average-spectrum}
\end{center}
\vspace{-0.75 cm}
\end{figure}

We obtained a rather good description of the source continuum spectrum with a simple power-law (PL) model, with photon index $\Gamma$ = 1.22 $\pm$ 0.04 and interstellar absorption $N_{\rm H}$ = (2.5 $\pm$ 0.1)$\times 10^{22}$ cm$^{-2}$; with this model we obtained \chisqnu/d.o.f. = 1.02/922. However, it was possible to significantly improve the spectral fit either by adding a partial covering fraction absorption (\textsc{tbpcf} in \texttt{XSPEC}) or by replacing the PL model with a cutoff power law (CPL). In both cases, the reduced chi-squared decreased to \chisqnu $\simeq$ 0.96. The observed source flux is \fx = (3.4$\pm$0.1)$\times 10^{-12}$ \flux. In Fig.~\ref{average-spectrum} we report the time-averaged source spectrum for the three EPIC cameras, together with the data-model residuals for the partially covered PL model (TBPCF$\times$PL). It shows that the source is strongly absorbed at E \lsim 1 keV.

%%%%%%%%%%%%%%%%%%%%%%%%%%%%%%%%%%%%%%%%%%%%%%%%%%%%%%%%%%%%%%%%%%%%%%%% 
%%%%%%%%%%%%%%%%%%%%%%%%%%%%%%%%%%%%%%%%%%%%%%%%%%%%%%%%%%%%%%%%%%%%%%%%
It was not possible to describe the source spectrum with other single-component models, though we obtained an equally good fit with a two-component model composed of a PL plus a thermal component, either a black body (BB) or a component due to collisionally ionised gas (\textsc{apec} model in \texttt{XSPEC}); in both cases the thermal component dominates the PL at E \lsim 1.4 keV. This additional component was significant at 99 \% c.l., and the F-test statistics value was \gsim 30, corresponding to a probability of \lsim 10$^{-13}$ that the additional thermal component is spurious. This proved that the fit improvement (compared with the simple PL model) was reliable. On the other hand, the addition of the thermal component to the TBPCF$\times$PL or the CPL model provided only a negligible improvement.

With all models we found weak positive residuals at E $\simeq$ 1.68 and 6.16 keV; moreover, in the case of the CPL model we also found hints of a rather wide emission feature at E = 0.98 keV. We found that, if these features are described with a Gaussian model, they are in some cases significant at 99 \% c.l. However, in order to assess the likelihood of theses features, we performed more in-depth investigations based on Monte Carlo simulations, using the \texttt{XSPEC} routine \textsc{simftes}. They revealed a high probability that the spectral data are consistent with emission models that do not include these features. Therefore, we ignored them in our analysis.

\begin{table*}[!t]
\caption{Best-fit parameters of the time-averaged EPIC spectrum of \cxo\ in the case of the four best-fitting continuum models.}\label{spectral_parameters}
\vspace{-0.5 cm}
\begin{center}
\begin{tabular}{cccccc} \hline \hline
Model                                   & -             & TBPCF$\times$PL 
      & CPL                     & PL+BB                 & PL+APEC       \\
Parameter                                       & Unit          & Value   
               & Value                                 & Value            
     & Value   \\ \hline
TBABS \nh                                                       & $\times10^{22}$  cm$^{-2}$      & 2.0$^{+0.3}_{-0.2}$   & 1.6 $\pm$ 0.2         & 4.1$\pm$0.4           & 4.0$^{+0.3}_{-0.4}$     \\
TBPCF \nh                                                       & $\times10^{22}$  cm$^{-2}$      & 6.7$^{+2.1}_{-1.7}$   & -     & -     & -     \\
TBPCF Covering Fraction & -     & 0.63$^{+0.07}_{-0.08}$        & -     & 
-       & -     \\
$\Gamma$                                        & -             & 1.66$^{+0.14}_{-0.13}$        & 0 $\pm$ 0.3             & 1.50$^{+0.09}_{-0.08}$ 
               & 1.49$\pm$0.08 \\
E$_{\rm cut}$                           & keV   & -             & 4.0 $^{+1.1}_{-0.8}$          & -                       & -     \\
Flux$_{\rm CPL/PL}$ (0.5-10 keV)$^{(a)}$        & $\times 10^{-12}$ \flux 
      & 6.4$^{+0.9}_{-0.6}$     & 3.88$^{+0.16}_{-0.15}$        & 5.5$^{+0.4}_{-0.3}$   & 5.5$\pm$0.3     \\
$kT_{\rm BB~or~APEC}$                   & eV    & -             & -       
              & 120$\pm$10              & 140$^{+30}_{-20}$     \\
$R_{\rm BB}^{(b)}$                              & km    & -             & 
-                       & 100$^{+80}_{-50}$     & -     \\
$N_{\rm APEC}$                          & cm$^{-5}$     & -     & -       
              & -                       & 0.6$^{+1.2}_{-0.4}$   \\
Flux$_{\rm BB~or~APEC}$ (0.5-10 keV)$^{(a)}$    & $\times 10^{-11}$ \flux 
      & -       & -                     & 2.9$^{+3.0}_{-1.5}$   & 22$^{+21}_{-14}$      \\ \hline
Flux$_{\rm BB~or~APEC}$/Flux$_{\rm TOT}$ (0.5-10 keV)   & -     & -     & 
-                       & 84.0 \%       & 97.6 \%       \\
Flux$_{\rm BB~or~APEC}$/Flux$_{\rm TOT}$ (0.01-100 keV) & -     & -     & 
-                       & 78.5 \%       & 99.6 \%       \\
Unabsorbed flux (0.5-10 keV)                    & $\times 10^{-11}$ \flux 
      & 0.64$^{+0.09}_{-0.06}$          & 0.39$^{+0.01}_{-0.02}$        & 
3.4$^{+3.0}_{-1.5}$   & 23$^{+20}_{-14}$        \\
Luminosity (0.5-10 keV)$^{(b)}$                 & $\times 10^{34}$ \lum & 
1.8$^{+0.3}_{-0.2}$     & 1.10$\pm$0.04 & 10$^{+8}_{-5}$        & 60$^{+60}_{-40}$      \\
\chisqnu/d.o.f.                         & -             & 0.95/920      & 
0.96/921                & 0.95/920              & 0.96/920      \\ \hline
\end{tabular}
\end{center}
Notes: $^{(a)}$ Corrected for absorption. $^{(b)}$ Assuming a source distance of $d$ = 5 kpc.
\end{table*}

In Table~\ref{spectral_parameters} we report the best-fitting parameters obtained for the four best-fit models. They show that the interstellar absorption is rather high since in all cases $N_{\rm H} \sim$ (2-4)$\times 10^{22}$ cm$^{-2}$. The source spectrum is rather hard since the photon index is $\simeq$ 0 in the case of the CPL model and $\simeq$ 1.5 in the other models. We note that, in the first case, the cutoff energy is rather low. In all models the flux of the PL component is a few 10$^{-12}$ \flux. For the thermal component, however, the estimated flux is much larger: one and two orders of magnitude higher for the BB and the APEC component, respectively. This component provides most of the unabsorbed flux for the combined PL and BB (PL+BB) model, and almost all the flux in the case of the combined PL and APEC (PL+APEC) model; in both cases the PL component only provides a marginal contribution at the high-energy end of the source spectrum. It is very probable that this result derives from the combination of the high interstellar absorption and the low temperature of the thermal component, which imply an overestimate of the thermal component's contribution to the unabsorbed flux. We also note that, for the PL+BB model, the normalisation of the BB component implies that the emitting region has a very large radius ($\simeq$ 100 km).

         %%%%%%%%%%%%%%%%%%%%%%%%%%%%%%%%%%%%%%%%%%%%%%%%%%%%%%%%%%%%%%%%%%%%
         \section{EPIC phase-resolved spectral analysis}
         \label{resolved-spectroscopy}
         %%%%%%%%%%%%%%%%%%%%%%%%%%%%%%%%%%%%%%%%%%%%%%%%%%%%%%%%%%%%%%%%%%%%

Although the folded light curves reported in Figs.~\ref{flc2E} and~\ref{flc4E} are characterised by CR maxima and minima that are almost aligned among the different energy ranges, they show that the pulse profile of \cxo\ is energy dependent. In fact, the HR is almost constant for about 60 \% of the pulse and shows two nearby peaks in the variable part. This suggests that the spectral properties of \cxo\ change along the pulse phase. In order to confirm or rule out this hypothesis, we conducted a phase-resolved spectral analysis in the two phase ranges, A ($\Delta\phi$ = 0.2-0.6) and B ($\Delta\phi$ = 0-0.2, 0.6-1), defined in Fig.~\ref{flc2E}, with the aim to analyse the pulse phases corresponding to a maximum or a minimum HR separately. Therefore, we selected two different spectra (A and B) for each EPIC camera. Then, since the count statistics of these spectra were rather low, we used the \texttt{SAS} task \textsc{epicspeccombine} to combine the three spectra of each phase range into a single spectrum. For both ranges we calculated the applicable response matrix and ancillary file.

\begin{figure}
\begin{center}
\includegraphics[width=8.5cm,angle=0]{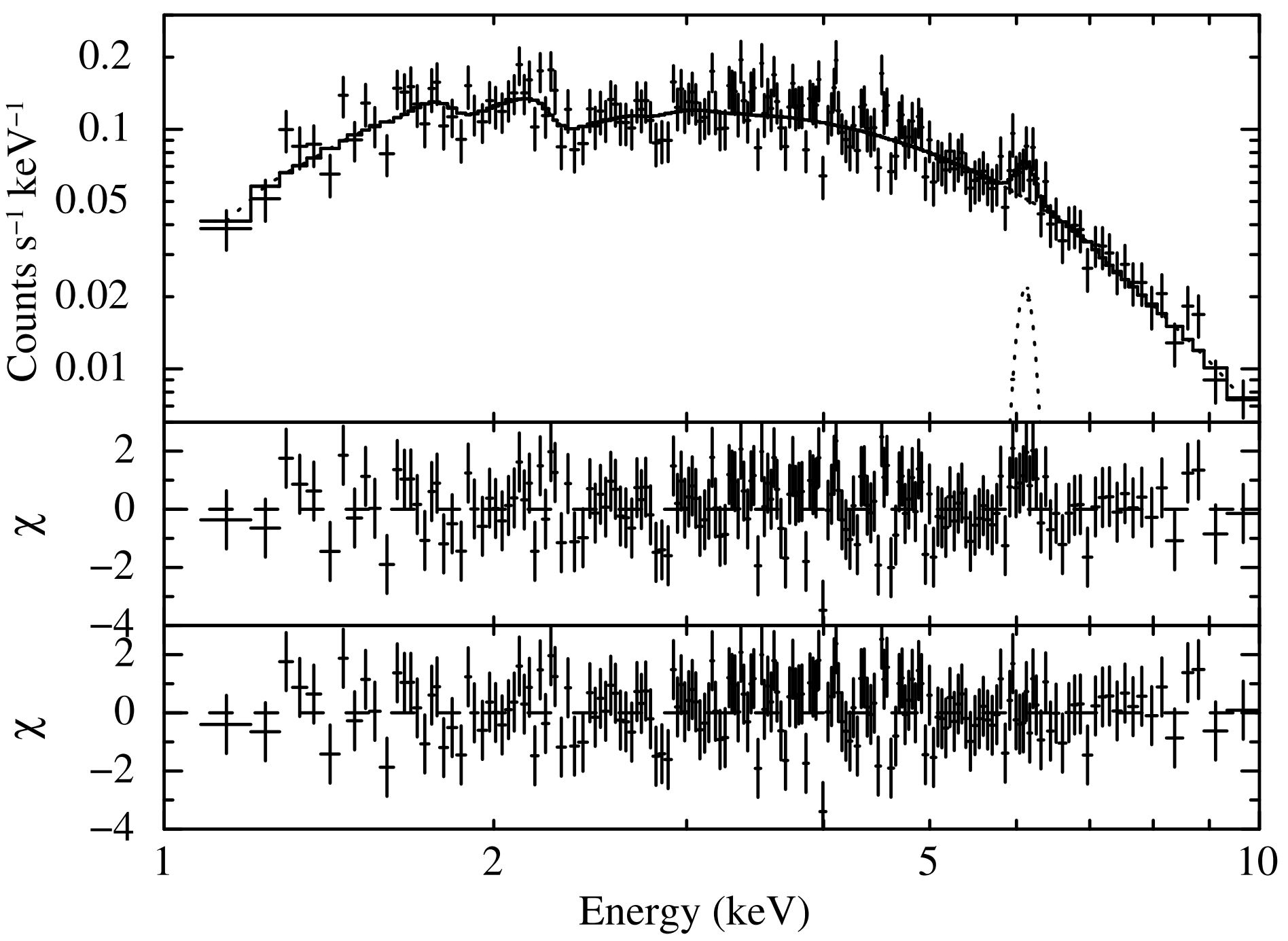}
\caption{EPIC spectrum of \cxo\ corresponding to phase range A. \textit{Upper panel}: Overlap of the spectrum with the best-fitting TBPCF$\times$(PL+Gaussian) emission model, where the Gaussian component is shown.
\textit{Middle panel}: Data-model residuals, in units of standard deviation ($\sigma$), in the case of the best-fitting TBPCF$\times$PL model.
\textit{Lower panel}: Data-model residuals, in units of standard deviation ($\sigma$), in the case of the best-fitting TBPCF$\times$(PL+Gaussian) model.}
\label{spectrum-hrmax}
\end{center}
\vspace{-0.75 cm}
\end{figure}

Our first task was to carry out an independent fit of the two spectra with the same emission models used for the time-averaged spectrum, with the aim of verifying if these models can also provide a good description of the phase-resolved spectra and, in this case, if the best-fitting values of the model parameters are compatible or inconsistent. In Table~\ref{phase-resolved-parameters} we report the results of this analysis. However, we noticed that spectrum A shows an evident data excess at E $\simeq$ 6.1 keV (Fig.~\ref{spectrum-hrmax}), while no similar feature is present in spectrum B. We verified that, regardless of the emission model for the continuum spectrum, this feature can be described with a Gaussian component, with energy E = 6.12$^{+0.08}_{-0.09}$ keV and intrinsic width $\sigma$ = 80$^{+110}_{-80}$ eV; its equivalent width (EW) is EW = 130$^{+90}_{-80}$ eV, and its flux is $7^{+4}_{-3}\times 10^{-6}$ \fph. Based on the estimated uncertainty of the line energy (at 90 \% c.l.), it is inconsistent with a neutral iron emission line. This component is significant at 99 \% c.l. with each of the continuum emission models, and its introduction implies a weak statistical improvement in the spectral fit, with a reduction in the \chisqnu\ value from 1.12 to 1.07 (in the case of the TBPCF$\times$PL emission model). Moreover, we verified with \textsc{simftest} that the probability that it is needless is below 1 \%  in all cases. However, since we cannot rule out that the performed analysis is affected by systematic errors, we think that the presence of this Gaussian component needs to be confirmed by better-quality data. For this reason, we did not introduce this component into the best-fit models of spectrum A, and we will not discuss it further.

\begin{table}[!t]
\caption{Best-fit parameters of EPIC spectra A and B of \cxo\ in the case of the four best-fitting continuum models.}\label{phase-resolved-parameters}
\vspace{-1 cm}
\begin{center}
\begin{tabular}{ccc} \hline \hline
Parameter                                                   & Spectrum A  
              & Spectrum B    \\ \hline
\multicolumn{3}{c}{TBPCF$\times$PL}    \\
TBABS \nh ($\times10^{22}$ cm$^{-2}$)   & 1.9$^{+0.6}_{-0.9}$       & 1.8$^{+0.3}_{-0.4}$       \\
TBPCF \nh ($\times10^{22}$ cm$^{-2}$)   & 6$\pm$2                       & 
6$\pm$2 \\
TBPCF Covering Fraction                 & 0.7$^{+0.2}_{-0.1}$       & 0.6$\pm$0.1       \\
$\Gamma$                                                    & 1.5$\pm$0.2 
                  & 1.6$^{+0.2}_{-0.1}$   \\
Flux$_{\rm PL}$ (0.5-10 keV)$^{(a)}$    & 6.7$^{+1.1}_{-0.8}$       & 5.4$^{+1.0}_{-0.6}$       \\
Luminosity (0.5-10 keV)$^{(b)}$                 & 1.9$^{+0.3}_{-0.2}$     
  & 1.5$^{+0.3}_{-0.1}$   \\
\chisqnu/d.o.f.                                     & 1.12/176            
      & 1.01/225                   \\ \hline
\multicolumn{3}{c}{CPL}    \\
TBABS \nh ($\times10^{22}$ cm$^{-2}$)   & 1.8$^{+0.5}_{-0.4}$       & 1.5$^{+0.2}_{-0.3}$       \\
$\Gamma$                                                    & -0.2$\pm$0.5                  & 0.2$^{+0.3}_{-0.4}$   \\
E$_{\rm cut}$ (keV)                         & 4$^{+2}_{-1}$               
      & 4$^{+2}_{-1}$          \\
Flux$_{\rm CPL}$ (0.5-10 keV)$^{(a)}$   & 4.3$^{+0.3}_{-0.2}$       & 3.5$\pm$0.2               \\
Luminosity (0.5-10 keV)$^{(b)}$                 & 1.23$\pm$0.08           
  & 0.99$^{+0.06}_{-0.05}$    \\
\chisqnu/d.o.f.                                     & 1.15/177            
      & 1.02/226                       \\ \hline
\multicolumn{3}{c}{PL+BB}    \\
TBABS \nh ($\times10^{22}$ cm$^{-2}$)   & 4.6$^{+0.8}_{-0.7}$       & 3.5$\pm$0.5               \\
$\Gamma$                                                    & 1.4$\pm$0.1 
              & 1.5$\pm$0.1           \\
Flux$_{\rm PL}$ (0.5-10 keV)$^{(a)}$    & 6.2$\pm$0.5               & 4.8$^{+0.4}_{-0.3}$   \\
$kT_{\rm BB}$ (eV)                          & 120$\pm$20                  
      & 120$^{+10}_{-20}$          \\
$R_{\rm BB}$ (km)$^{(c)}$               & 110$^{+150}_{-70}$        & 80$^{+90}_{-50}$      \\
Flux$_{\rm BB}$ (0.5-10 keV)$^{(a)}$    & 40$^{+90}_{-30}$          & 20$^{+20}_{-10}$      \\
Flux$_{\rm BB}$/Flux$_{\rm TOT}$ (0.5-10 keV)   & 85.8 \%               & 
77.0 \%             \\
Flux$_{\rm BB}$/Flux$_{\rm TOT}$ (0.01-100 keV) & 76.9 \%               & 
70.7 \%             \\
Unabsorbed flux (0.5-10 keV)$^{(a)}$    & 50$^{+90}_{-30}$              & 
20$^{+30}_{-10}$            \\
Luminosity (0.5-10 keV)$^{(b)}$                 & 14$^{+25}_{-9}$         
  & 6$^{+7}_{-3}$         \\
\chisqnu/d.o.f.                                     & 1.11/176            
      & 1.01/225                   \\ \hline
\multicolumn{3}{c}{PL+APEC}    \\
TBABS \nh ($\times10^{22}$ cm$^{-2}$)   & 4.4$\pm$0.7               & 3.4$^{+0.4}_{-0.5}$   \\
$\Gamma$                                                    & 1.4$^{+0.1}_{-0.2}$       & 1.5$\pm$0.1           \\
Flux$_{\rm PL}$ (0.5-10 keV)$^{(a)}$    & 6.0$^{+0.6}_{-0.4}$       & 4.8$\pm$0.3           \\
$kT_{\rm APEC}$ (eV)                    & 140$^{+40}_{-30}$               
  & 140$^{+40}_{-20}$      \\
$N_{\rm APEC}$ (cm$^{-5}$)                              & 1.0$^{+0.7}_{-0.8}$           & 0.30$^{+0.23}_{-0.25}$  \\
Flux$_{\rm APEC}$ (0.5-10 keV)$^{(a)}$  & 300$^{+900}_{-200}$       & 120$^{+210}_{-90}$        \\
Flux$_{\rm APEC}$/Flux$_{\rm TOT}$ (0.5-10 keV)     & 98.2 \%       & 95.9 \%                  \\
Flux$_{\rm APEC}$/Flux$_{\rm TOT}$ (0.01-100 keV)       & 99.6 \%       & 
99.3 \%             \\
Unabsorbed flux (0.5-10 keV)$^{(a)}$    & 300$^{+1000}_{-200}$      & 120$^{+210}_{-90}$        \\
Luminosity (0.5-10 keV)$^{(b)}$                 & 100$^{+290}_{-70}$      
  & 30$^{+60}_{-20}$         \\
\chisqnu/d.o.f.                                     & 1.11/176            
      & 1.02/225                   \\ \hline
\end{tabular}
\end{center}
Notes: $^{(a)}$ Corrected for absorption, $\times 10^{-12}$ \flux. $^{(b)}$ $\times 10^{34}$ \lum, assuming a source distance of $d$ = 5 kpc. $^{(c)}$ Assuming a source distance of $d$ = 5 kpc.
\end{table}

We verified that the source flux decreases from (3.8$\pm$0.1)$\times 10^{-12}$ for spectrum A to (3.0$\pm$0.1)$\times 10^{-12}$ \flux\ for spectrum B, with a reduction of $\simeq$ 20 \%. Taking into account the estimated uncertainties, the best-fit values of most parameters are consistent between the two spectra as well as with the results obtained for the time-averaged spectrum. The unique exception is the absorbing hydrogen column density in the cases of the PL+BB and PL+APEC models, which is significantly lower for spectrum B than for spectrum A. In all models the flux of the non-thermal component decreases by $\simeq$ 20 \% from spectrum A to spectrum B. The flux reduction is much larger for the thermal components ($\simeq$ 60 \% for the BB and $\simeq$ 70 \% for the APEC), but, due to the large uncertainties, the two spectra's fluxes are not clearly inconsistent.

These results provide no clear evidence of a spectral variation between the two phase ranges; in the case of the PL-plus-thermal-component model, they only suggest a variation in the relative contribution of the two components. In order to further investigate the source spectral variability, we carried out a simultaneous fit of the two spectra using the same best-fit models reported in Table~\ref{phase-resolved-parameters}.

In the case of the two models without the thermal component, we tried to fit both spectra by only allowing independent values for the normalisations of the PL or CPL components, while for all the other spectral parameters we forced a common value for the two spectra. This approach resulted in an unsuccessful fit since in both cases the null hypothesis probability (NHP) of the best-fit model was $< 2\times 10^{-3}$. This suggests that the variation in the normalisation in the emission components cannot solely account for the difference between the two spectra. A variation in the absorption, photon index, and/or cutoff energy is also needed. We also tried to fit the two spectra with a common PL or CPL component while leaving the absorption components free to vary independently. In the first case we only obtained a fit improvement (\chisqnu/d.o.f. = 1.14/401 with NHP $\simeq$ 0.03) with a very large variation in the TBPCF parameters. In fact, \nh\ increased from 5.8$^{+1.5}_{-1.2}$ to 17$^{+8}_{-5}$ $\times10^{22}$ cm$^{-2}$, while the covering fraction decreased from 0.8$\pm$0.1 to 0.55$^{+0.07}_{-0.10}$. In the second case, on the other hand, a common CPL component was rejected by the data.

We also used the same approach in the two cases of the PL-plus-thermal-component model, where we also imposed a common value for the temperature of the thermal component. In both cases we obtained a best-fit solution with NHP $\simeq$ 0.06 and 0.03 for the PL+BB and PL+APEC models, respectively. Therefore, in these cases we cannot rule out that the spectral variability is simply due to a variation in the relative contribution of the different components instead of to an intrinsic change in their shapes. We verified that, in both cases, a common PL component is rejected by the data since NHP \lsim $10^{-3}$. Finally, we also investigated the possibility that the thermal component does not vary between the two phase ranges. Therefore, we repeated the simultaneous fit of the two spectra by imposing a common value for both their temperatures and normalisations. This possibility was rejected by the data (NHP $< 4\times 10^{-3}$) if we only allowed independent values for the PL normalisation, while it was clearly acceptable (NHP $>$ 0.1) if at least one of the other parameters (either the absorption or the photon index) was allowed to have independent values between the two spectra. These results suggest that the spectral data are consistent with a constant thermal component.
%%%%%%%%%%%%%%%%%%%%%%%%%%%%%%%%%%%%%%%%%%%%%%%%%%%%%%%%%%%%%%%%%%%%%%%% 
%%%%%%%%%%%%%%%%%%%%%%%%%%%%%%%%%%%%%%%%%%%%%%%%%%%%%%%%%%%%%%%%%%%%%%%%

         %%%%%%%%%%%%%%%%%%%%%%%%%%%%%%%%%%%%%%%%%%%%%%%%%%%%%%%%%%%%%%%%%%%%
         \section{Discussion}
         \label{discussion}
         %%%%%%%%%%%%%%%%%%%%%%%%%%%%%%%%%%%%%%%%%%%%%%%%%%%%%%%%%%%%%%%%%%%%

\subsection{The optical counterpart and the source distance}

The spectral properties of the optical counterpart have been investigated by \citet{Esposito2013}, \citet{Lutovinov2013}, and \citet{Masetti2013}, while the near-infrared spectrum has been reported by \citet{Rodes-Roca2018}. These authors all favoured a classification of the X-ray source as a BeXB located at about 4-5 kpc. 
 
In particular, the high resolution optical spectrum investigated by \citet{Esposito2013} led to the classification of the counterpart as a B0-B1 type star, most likely a B1V star located at a distance of 4-5 kpc, although a B III donor could not be excluded. A luminosity class of I or II was also discussed, implying a distance around 14–19 kpc and 9–12 kpc, respectively. While a very distant supergiant star was excluded by the position of the Galaxy edge at about 10 kpc, a B1II companion could not be completely ruled out, despite being judged unlikely, because of the lack of some typical spectral features.
  
Recently, a parallax measurement became available thanks to the Gaia satellite (albeit with a low signal-to-noise ratio, $p = 0.053 \pm 0.037$\,mas), pointing to significantly larger values for the distance to the source ($8.1^{+2.2}_{-1.6}$\,kpc; \citealt{Bailer-Jones2018}). However, we note that the Gaia distance does not call into doubt the BeXB classification. A viable possibility to reconcile the optical spectrum with a larger distance is to assume a giant star, which would be located at about 5-9 kpc. Indeed, assuming a B1 III star (with $(B-V)_0=-0.21$ and $M_V=-3.71$; \citealt{Wegner1994, Wegner2006}) from the observed colour in the range $(B-V)_{\rm obs} = 1.4$--1.8 (Table\,5 in Esposito et al. 2013), we calculate an excess colour of $E(B-V)=1.6$--2, leading to a source distance in the range $\sim$5--9 kpc, which is consistent with the Gaia result. A giant star was also favoured by \citealt{Masetti2013}, although they analysed an optical spectrum with a lower resolution than that reported by \citealt{Esposito2013}.

\subsection{Timing and spectral properties}
\begin{figure}
\centering
\resizebox{\hsize}{!}{\includegraphics[angle=0]{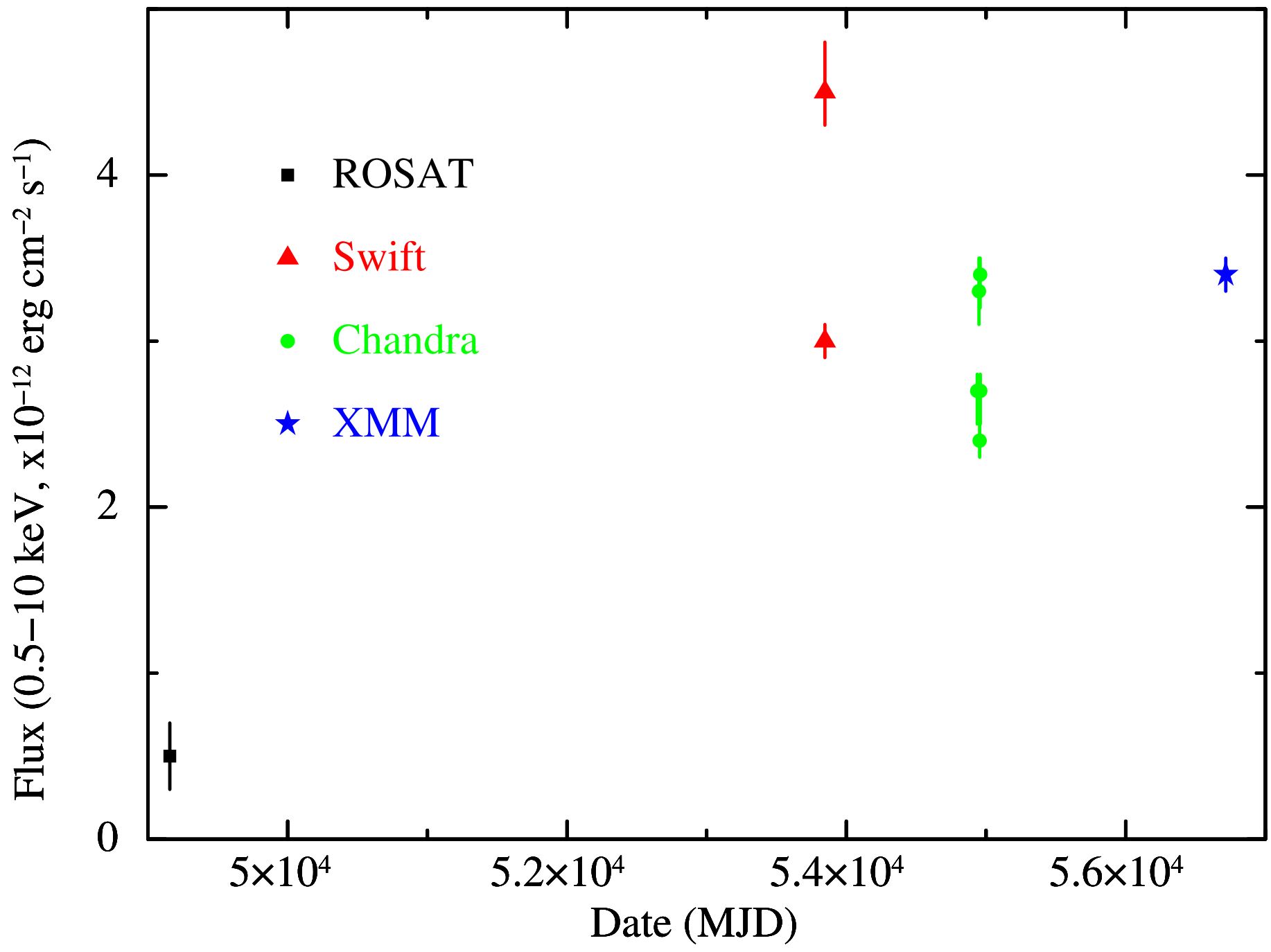}}
\caption{Long-term evolution of the absorbed flux of \cxo, as measured with \emph{ROSAT}, \emph{Swift}, \emph{Chandra}, and \XMM\ (updated from \citealt{Esposito2013}).}
\label{flux}
\end{figure}

The \XMM\ observation of \cxo\ performed in 2014 detected the source at a flux level of \fx = (3.4$\pm$0.1)$\times 10^{-12}$ \flux. In Fig.~\ref{flux} we report this flux together with all the flux values measured in the previous observations. It shows that, with the only exception of the first \ROSAT\ observation in 1992, the source flux is rather stable, with variations of $\pm$ 30 \% at most. Regardless of the emission model adopted to describe the source spectrum, and assuming a source distance of 5 kpc, the unabsorbed flux implies a luminosity of \lx $< 10^{36}$ \lum. Therefore, compared to most of the BeXBs \citep{HaberlSturm16,Tsygankov+17}, the X-ray emission of \cxo\ is characterised by reduced variability and relatively low luminosity.

\begin{figure}
\centering
\resizebox{\hsize}{!}{\includegraphics[angle=0]{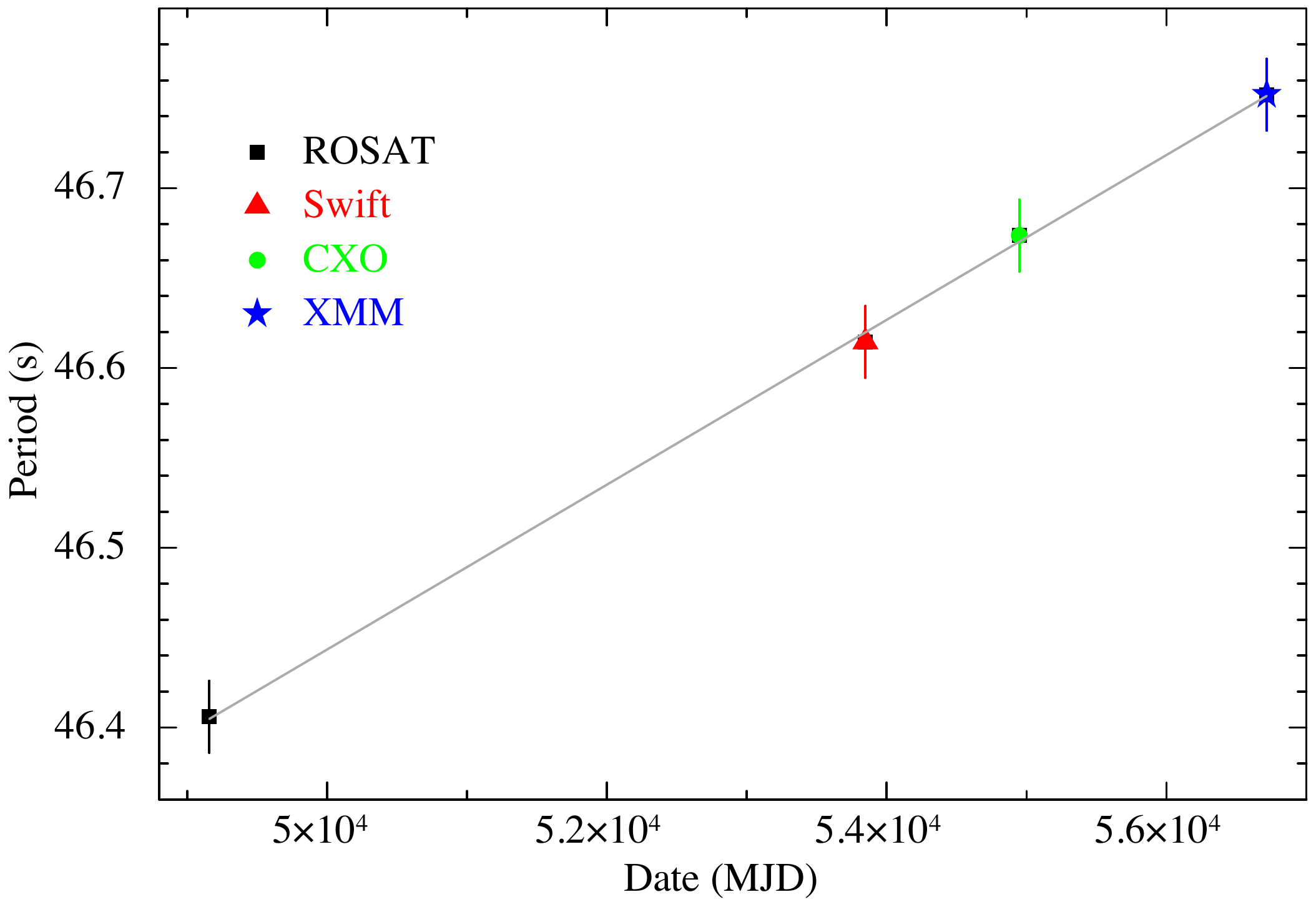}}
\caption{Long-term spin down of \cxo\ from \emph{ROSAT}, \emph{Swift}, \emph{Chandra}, and \XMM\ measurements (updated from \citealt{Esposito2013}). We attributed an error of 0.02\,s to each point, reflecting the uncertainty due to the orbital motion (see \citealt{Esposito2013} for details). The grey line indicates the best-fitting period derivative, $\dot{P}\approx5.3\times10^{-10}$\,s\,s$^{-1}$.}
\label{pdot}
\end{figure}

The pulse period measured with \XMM\ in February 2014 is $P$ = 46.753(3) s. In Fig.~\ref{pdot} we report all the pulse period values measured to date; for all of them we have assumed a systematic error of 0.02 s, in agreement with the uncertainty due to the orbital motion of the pulsar \citep{Esposito2013}. The figure shows that the \XMM\ value is the highest one and that the period variation is much larger than the variation due to the orbital motion. Moreover, the \XMM\ value is fully consistent with a constant pulsar spin down at an average rate of $\pdot = 5.3\times10^{-10}$ s s$^{-1}$, in agreement with the findings of \citet{Esposito2013}.

The pulse profile obtained with \XMM\ confirms the three-peak structure already found with \Chandra\ and, in addition, reveals an energy dependence. The three peaks are more evident at low energies, and the HR is anti-correlated with the CR. This HR variation could be related to a variability in the material column density along the line of sight since the phase-resolved spectroscopy shows that the photon absorption increases when the CR decreases (Table~\ref{phase-resolved-parameters}). The PF is $\simeq$ 40-45 \%, slightly increasing with the photon energy. These results prove the spectral variability along the spin phase, which was only suggested by the \Chandra\ data.

As in the case of the \Swift\ and \Chandra\ observations, the time-averaged EPIC spectra of \cxo\ can be described reasonably well with an absorbed PL emission model, although both the interstellar absorption \nh\ and the photon index $\Gamma$ are slightly higher than in the previous observations. However, we obtained a better spectral fit using four different spectral models. For two of them we only considered a non-thermal component (either a partially covered PL or a CPL model), while for the other two we introduced an additional thermal component (either a BB or an APEC model) to the original PL component. In all cases we obtained the same fit quality, so it was not possible to prefer or reject any emission model on a statistical basis.

For the two models with only a non-thermal component, the estimated interstellar absorption was \nh\ $\simeq 2 \times 10^{22}$ cm$^{-2}$, a value almost equal to the result obtained with the \ROSAT, \Swift, and \Chandra\ data. We note that this value is a factor of $\simeq$ 2 higher than the estimated Galactic value in the source direction (\nh$_{,Gal} = 9\times10^{-21}$ cm$^{-2}$). Moreover, the \nh\ value is even higher for the other two emission models. This implies that $\sim$ 50 \% of the total absorption is due to a local component. With the only exception of the CPL model, in all cases the PL photon index is $\Gamma \simeq$ 1.5. This result is rather similar to those obtained in the previous observations, where $\Gamma$ is always in the range 1.2-1.5. In the case of the partially absorbed PL model, the TBPCF component represents an inhomogeneous absorber medium. It can be reasonably ascribed to the Be, clumpy, polar wind crossing the line of sight to the X-ray source.  

In the other two emission models we investigated the presence of an additional thermal component. Indeed, it is well known that several BeXBs (either persistent with low luminosity or transient with high luminosity) are characterised by this type of feature \citep{LaPalombara+12,LaPalombara+18}. In the case of \cxo, we found that, for both the BB and the APEC model, the estimated temperature is rather low ($T$ = 120-140 K). The estimated unabsorbed flux of the thermal component largely dominates the total source flux. In the case of the PL+APEC model in particular, the PL contribution to the total flux is almost negligible. Moreover, although in both models the parameters of the thermal component are characterised by large uncertainties, they suggest that the size of the thermal emission region is significantly larger than the NS size. For all these reasons, we disfavour the presence of such a thermal component in the X-ray emission from the pulsar.

In the phase-resolved spectral analysis, we considered the two spectra corresponding, respectively, to the hard (A) and soft (B) parts of the pulsation separately. We first performed an independent fit of the two spectra using the same four emission models adopted for the time-averaged spectral analysis. We found that the source total flux decreases by $\simeq$ 20 \% from A to B and that the same relative decrease occurs for the non-thermal component; the flux reduction of the thermal component (either BB or APEC) is much higher (60-70 \%), but the flux estimates are affected by large uncertainties. As such, it is not possible to claim a firm flux variability. Since the independent fit of the two spectra revealed a flux variation but no evidence of variability for the spectral parameters, we also performed a simultaneous fit of the two spectra. Doing so, we were able to determine that a simple flux variation cannot explain the source variability. In fact, for the non-thermal models it was not possible to obtain an acceptable fit when only the normalisations were allowed to vary. However, for the models with a thermal component, the spectral variability can be described with a change in the relative contribution of the two components (thermal and non-thermal); on the other hand, a constant thermal component is acceptable only if the absorption and/or the photon index vary. Nevertheless, we note that this analysis only proves the spectral variability between phase ranges A and B. Due to the uncertainties regarding the presence of the thermal component, it is not possible to provide any further interpretation.

Finally, we searched the $INTEGRAL$ archive (\cxo\ is associated with the hard X-ray source IGR~J22534+6243; \citealt{Landi2012,Krivonos2012}) for eventual outbursts or bright flares. However, we found no detections on a timescale of $\sim$2 ks (i.e. the duration of each single \emph{INTEGRAL} pointed observations) between October 2002 and January 1, 2016, in any of the following energy ranges: 18-50 keV, 22-50 keV, 50-100 keV, and 100-150 keV.

         %%%%%%%%%%%%%%%%%%%%%%%%%%%%%%%%%%%%%%%%%%%%%%%%%%%%%%%%%%%%%%%%%%%%
         \section{Conclusions}
         \label{conclusions}
         %%%%%%%%%%%%%%%%%%%%%%%%%%%%%%%%%%%%%%%%%%%%%%%%%%%%%%%%%%%%%%%%%%%%

The results provided by the \XMM\ observation of \cxo\ in 2014 confirm that this BeXB is characterised by a low luminosity (\lx $\sim 10^{34-35}$ \lum) and limited variability (within $\pm$ 30 \% at most). The measured pulse period, $P$ = 46.753(3) s, compared with the previous measurements, is fully consistent with a constant pulsar spin down at an average rate of $\pdot = 5.3\times10^{-10}$ s s$^{-1}$. These findings are coherent with a wind-accretion scenario of the pulsating NS from the companion Be star.

The properties of the thermal emission component observed in several X Per-like sources are not contradicted by the spectral characterisation we obtained for \cxo. Therefore, the results provided by the \XMM\ observation provide more evidence of the similarity between \cxo\ and X Per-like sources, thus confirming what was suggested by \citet{Esposito2013}: in the class of persistent BeXBs, \cxo\ is the source with the shortest pulse period.

\begin{acknowledgements}
GLI acknowledges funding from ASI-INAF agreements I/037/12/0 and 2017-14-H.O. PE and GLI also acknowledge financial support from the Italian MIUR PRIN grant 2017LJ39LM. This research has made use of the local INTEGRAL archive (ANITA, A New InTegral Archive) developed at INAF-IASF Milano  (https://www.iasf-milano.inaf.it/ANITA).
\end{acknowledgements}

\bibliographystyle{aa}
\bibliography{biblio}

\end{document}